\documentclass[10pt, preprint]{aastex}
\usepackage{graphicx}

\begin{document}
\title{Interaction of Jets with the ISM of Radio Galaxies}

\shorttitle{Interaction of Jets with ISM}        % if too long for running head
\shortauthors{Sutherland et al.}

\author{Ralph S. Sutherland and Geoffrey V. Bicknell}
\affil{Research School of Astronomy \& Astrophysics, Australian National University, Cotter Rd., Weston ACT, Australia 2611}

\begin{abstract}
We present three dimensional simulations of the interaction of a light hypersonic jet with an inhomogeneous 
thermal and 
turbulently supported disk in an elliptical galaxy. 
These simulations are applicable to the GPS/CSS phase of 
some 
extragalactic radio sources.  

We identify four generic phases in the evolution of such a jet with the interstellar medium. The first is a `flood and channel'' phase, dominated by complex jet interactions with the dense cloudy medium close to the nucleus. This is characterized by high pressure jet gas finding changing weak points in the ISM and flowing through channels that form and reform over time.   A spherical, energy driven, bubble phase ensues, wherein the bubble is larger than the disk scale, but the jet remains fully disrupted close to the nucleus, so  that the jet flux is thermalised and generates a smooth isotropic energy-driven bubble.   In the subsequent, rapid, jet break--out phase the jet breaks free of the last obstructing dense clouds, becomes collimated and pierces the more or less spherical bubble. In the final classical phase, the jet propagates in a momentum-dominated fashion similar to jets in single component hot haloes, leading to the classical jet -- cocoon -- bow-shock structure. 
\end{abstract}


\section{Introduction}
\label{intro}

In this paper we present a high resolution simulation of a radio jet interacting with an inhomogeneous interstellar medium in the form of a turbulently supported disk which extends our previous work in several different directions: (1) The simulation is three dimensional with a resolution of 512 cells in the (cartesian) coordinate directions and a spatial scale of 2 parsecs per cell. (2) The density structure of the warm disk is described by a log-normal distribution, typical of the warm interstellar medium in a number of different environments. (3) The initial distribution of the warm medium is that of a thick almost Keplerian disk supported by a combination of thermal pressure and a, dominant, supersonic velocity dispersion. This type of distribution is one of several that may be contemplated and is supported by the observation of such disks in M87 \citep{dopita97a} and NGC~7052 \citep{vandermarel98a}.  (4) The soft X-ray emissivity and surface brightness of the thermal plasma is calculated, in conjunction with the radio surface brightness image. These synthetic images provide valuable insights for the interpretation of observational data.  

This simulation illustrates four important phases in the evolution of a young radio galaxy: (1) An initial ``flood and channel" phase wherein the radio source is beginning to be established, but is still interacting strongly with the disk material. (2) A quasi-spherical jet-driven bubble phase where the jet is fully disrupted.  (3) a 
rapid jet establishment phase
where the last obstruction is ablated away and the jet reforms and crosses the spherical bubble.  (4) A classical jet -- cocoon -- bow-shock phase familiar from previous studies of radio sources in a single component, hot ISM.  We also determine the X-ray morphology associated with the interaction of the radio source with the disk and show that the luminosity from the disk may persist well after the initial flood and channel phase
. 

\section{Model Parameters}
\label{sec:1}
\subsection{Host galaxy potential}

Elliptical galaxies contain both baryonic and dark matter. The former dominates on scales close to the core and the latter dominates at radii $\sim 10 kpc$.
Since jet simulations frequently extend over such scales,
 we have constructed a family of potentials which we use to represent a combined, self-consistent  distribution of baryonic and dark matter. In the simulation presented here, the scale of the simulation is such that the potential is dominated by the luminous component close to the core. The potential is based on isothermal distributions of baryonic and dark matter, described in terms of isotropic distribution functions.

Let $\sigma_D$ and $r_D$ be the velocity dispersion and core radius of the dark matter distribution. Normalising the potential by $\sigma_D^2$ and the radius by $r_D$, the dimensionless version of Poisson's equation is:
\begin{equation}
\frac {d^2 \psi}{dr^{\prime 2}} + \frac{2}{r^\prime} \frac {d \psi}{dr^\prime} =
9 \, \left[ \exp(-\psi) + \varrho  \exp \left( -\kappa^2 \psi \right)\right] \, .
\label{e:poisson}
\end{equation}
It is convenient to take  the ratio of dark to luminous velocity dispersions, $\kappa$, and the the ratio of core radii, $\lambda$, as defining parameters of the potential and mass distribution.
The central densities are given by:
\begin{equation}
\rho_{\rm D,0} = \frac {9 \sigma_{\rm D}^2}{4 \pi G r_{\rm D}^2} \, ,
\qquad 
\rho_{\rm B,0} = \frac {9 \sigma_{\rm B}^2}{4 \pi G r_{\rm B}^2} \, .
\end{equation}
We numerically integrate (\ref{e:poisson}) for chosen parameters, tabulate the results and interpolate when required.  In this paper we use $\kappa = 2$, and $\lambda = 10$,  referring to the potential as $\psi_{2,10}$.  The central density is dominated by baryonic material over dark matter by a ratio of $25:1$.  

 We adopted this isothermal formalism for the galactic potential for two reasons.  First, it seems to be supported by observations of at least some classes of galaxies, ( c.f. \citet{cowsik2007},  \citet{kdenaray2006}  , \citet{deblok2005}) although perhaps not all (e.g. \citet{schmidt2007} ).  
Secondly, the well known Navarro, Frenk \& White, (NFW), \citep{NFW1997}, profile arises in an empirical manner from smooth particle cosmological simulations, whereas the isothermal formalism arises from simple physical considerations \citep{king1966}.  In the absence of a definitive proof either way, we choose to follow the isothermal path.   We note that with a coupled baryonic/dark matter potential, the region between the baryonic core radius (350~pc) and the outer dark matter core radius (3.5~kpc) does produce intermediate density radial power--law indices in the range; $0 < \alpha < -1$, before turning down to steeper slopes beyond the dark matter core radius (well outside our present simulation).   If a galaxy with such a double potential were observed with insufficient resolution to resolve the inner 300~pc, then the observed distribution inside the dark matter core radius would be observed to be steeper than a single isothermal distribution, perhaps being consistent with a more `cuspy' NFW distribution. (See \cite{sutherland2007} for more details).

\subsection{The Hot Galactic Atmosphere}

The tenuous, hot interstellar medium in our simulations is isothermal. For gas temperature $T$ the dark matter virial temperature is $T_* = \bar{m}  \sigma_{\rm D}^2 / k$, where $ \bar{m}  = \mu m_u \approx 1.035\times 10^{-24}$~g, is the mean particle mass.  Assuming hydrostatic equilibrium, the particle number density, $n_{\rm h}$ of the hot gas is 
${n_{\rm h}}/{n_{\rm h,0}} = \exp \left[ -{T_*}/{T} \psi  \right]$.

\subsection{ The non-uniform, fractal, warm interstellar medium}
\label{s:warm_ISM}

To establish a non-uniform interstellar medium we utilise an obvious analogy with a turbulent medium.  Rather than attempt to review the vast topic of turbulence here, we refer the reader to the recent astrophysically oriented reviews of \citep{elmegreen04}, and \citep{scalo04} as a starting point and then refer simply to some specific results that we use below to arrive at reasonable parameters for our non-uniform medium. 

\subsubsection{Log-normal density distribution}
\label{s:lognormal}

We use a log--normal distribution to describe the single point statistics of the density field of our nonuniform ISM.  The log--normal distribution is a skewed continuous probability distribution, for an independent variable $x \geqslant 0$.  Unlike the normal distribution, it has a non-zero skewness, variable kurtosis/flatness, and in general the mode $<$ median $<$ mean.   The log-normal distribution appears to be a nearly universal property of isothermal turbulent media, both experimentally, numerically and analytically ( {\em e.g.} \citet{nordlund99a} , see also \citet{warhaft00} ).  It is also at least intuitively encouraging  that is corresponds the to the limiting distribution for the product of multiplicative random increments, in the same way that the normal distribution plays that role for additive random increments in the {\em central limit theorem}. It is thus compatible at least conceptually with a generic cascading process consisting of repeated folding and stretching.

The probability density function with parameters $s$ and $m$ (the standard deviation and mean of the corresponding normal distribution), for the log--normal ISM density $\rho$ is,
$$P(\rho) ={1}/({s \sqrt{2 \pi} \, \rho}) \exp \left \lbrack {-(\ln \rho - m)}/{2s^2}\right \rbrack \, .$$

We adopt a mean $$\mu =  \exp[ m + s^2/2] = 1.0 \, ,$$ and $$\sigma^2 =  \mu^2 \, (\exp[ s^2]-1) = 5.0\, ,$$ as our standard log--normal distribution, compatible with the favored ranges in \citet{fischera03} and \citet{fischera04} in their star burst galaxy reddening/exctinction models.  The  variance determines the concentration of mass is into dense cores, or conversely the volume occupied by voids.   With these parameters, densities below the mean, $\mu$, comprise one quarter of the mass, and  three quarters of the volume; the mean is approximately 20 times the mode.

\subsubsection{Powerlaw density structure}

The isotropic power spectrum of the fourier transform of the density, $F(k)$ is $D(k) =  \int  k^2 F(k) F^*(k) \> d \Omega$,
where the angular integrals are over the relevant angular variables in Fourier space.
If $D(k) \propto k^{-\beta}$, and if $\beta = 5/3$, the spectrum is often referred to as Kolmogorov turbulence. A scalar tracer (density) of the turbulent field also shows the Kolmogorov structure index \citep{warhaft00}; hence we generate a density distribution with the Kolmogorov index for our inhomogeneous ISM.

\subsubsection{Equilibrium turbulent disks}

Many radio galaxies exhibit a turbulent disk of gas in the central regions. We establish equilibrium turbulent disks in the appropriate potential by first determining the ensemble average distribution of the gas and then realizing a single instance of the ensemble by multiplying by a log-normal power-law structure 
as described above  with unit mean, variance $\sigma^2 = 5.0$ and power-law index $\beta = 5/3$.  

Following, \citet{strickland00a} we adopt the ansatz of an {\em almost} Keplerian disk. We find:
\begin{equation}
\frac {\bar \rho_{r,z}}{\bar\rho_{0,0}} = \exp \left[  -\frac {\sigma_D^2}{\sigma_g^2}
\{ \psi_{r,z} - e^2 \psi_{r,0}- (1-e^2) \psi_{0,0}  \} \right] \, .
\label{e:warm_gas}
\end{equation} 
where $\sigma_g$ is the turbulent velocity dispersion of the gas.
The main differences from the development of a similar equation by \citet{strickland00a} is the 
(logically required)
 lack of dependence of $e_K$ on $z$ and the formal introduction of a turbulent velocity. The latter avoids the difficulty of prescribing an unphysically large temperature in order to achieve a reasonable disk scale height.

In summary we use the double isothermal potential $\psi_{\kappa, \lambda}$with $\kappa = 2$, $\lambda = 10$, 
scaled by $r_D = 3.5$~kpc, and $\sigma_D = 400$~km/s, and a rotation parameter $e_{\rm K} = 0.93$, plus a velocity dispersion of the warm gas of $\sigma_g = 40$~km/s for the disk. 

\section{Jet Parameters}

We use a non-relativistic code for these simulations. The use of such a code  to simulate phenomena that involve relativistic flow in parts of the grid, is not ideal. Nevertheless, a large part of the flow field is non-relativistic and is driven by the energy or momentum flux provided by the jet. Hence, we establish jet parameters in such a way that the energy flux of the non-relativistic jet corresponds to the jet energy flux of a given relativistic jet using relationships derived by \citet{komissarov96a}.

The density ratio, $\eta$ and the Mach number, $M_{\rm nr}$ of a non-relativistic jet corresponding to a jet with Lorentz factor $\Gamma$ and density parameter $\chi = \rho c^2 /4 p$ is described by the following equations. The parameters $\xi$ and $T_{\rm ism}$ are the ratio of jet to external pressures  and the (hot) interstellar medium temperature respectively.
\begin{eqnarray}
\eta &=& \frac {2 \gamma}{\gamma-1} \xi 
\left( \frac {kT_{\rm ism}}{\mu m c^2} \right) \Gamma^2 
\left[ 1 + \frac {\Gamma}{\Gamma+1} \chi \right] \, .
\label{e:eta}\\ 
M_{\rm nr}^2 &=&\frac {2}{\gamma-1} \left[ 1 + \frac {\Gamma}{\Gamma+1} \chi \right]
\left( \Gamma^2 - 1 \right) \, .
\label{e:M_nr}
\end{eqnarray}
Note that the low value of $kT_{\rm ism}/\bar{m}c^2 \sim 10^{-6}$ guarantees a light non-relativistic jet ($\eta \ll 1$) and that the non-relativistic Mach number effectively corresponds to the Lorentz factor.

In our simulations we take the gas to have the ideal adiabatic index, $\gamma=5/3$, since this represents the external medium the most accurately and we are mainly interested in the effect that the jets have on the external clumpy medium.

\subsection{Resolution and Scaling}

We have selected a jet for which the instantaneous hot spot
$\beta_{\rm hs} \sim 0.05 - 0.1$ and the initial jet diameter is $ 20 \> \rm pc$. These parameters, together with a central ISM  $p/k = 10^6 \> \rm cm^{-3} \> K$ places the jet at the low end of FR2 power $\approx 3 \times 10^{43} \> \rm ergs \> s^{-1}$. This choice of parameters produces a jet which initially interacts strongly with the ambient ISM but whose morphology at later times is similar to a classical FR2 radio source.

With adiabatic simulations, the choice of the spatial, velocity and density scales is arbitrary. However, the introduction of cooling (in the thermal gas) restricts the allowable scaling to a one-parameter set.  Here a given simulation defines a one-parameter family of simulations where the scaling parameters satisfy the following constraints:
\begin{equation}
\begin{array}{r c l r c l }
x_0 &=& \hbox{arbitrary} ,  & v_0 &=& \hbox{fixed} \, , \\
 t_0 &=&{x_0}/{v_0}\, ,        & \rho_0 &\propto & x_0^{-1} \, ,\\
 p_0 &=& \rho_0 v_0^2 \, ,          & {kT_0}/{\bar{m}} &=& v_0^2\, , \\
\rho_{*,0} &\propto & x_0^{-2}\, ,  & M_{*,0}  &\propto & x_0\, .
\end{array}
\end{equation}
with the proviso that $x_0$ cannot increase to the extent that the gas density exceeds the density of gravitating matter.
Notwithstanding this restricted one-parameter scaling, the allowable set of physical models allowed by this scaling describes an interesting variety of different physical situations. (See Sutherland \& Bicknell 2007 for further details)

Three simulations were performed, models {\bf A}, {\bf B}, and {\bf C}. The 'Standard Model' {\bf A}) is presented in the greatest detail. Models {\bf B} and {\bf C} are used to compare the effect of the density and the distribution (smooth or fractal) of the warm gas.  Table~\ref{t:Model_A_pars} summarizes all the parameters used in Model~{\bf A}, for the jet, the potential and the interstellar medium respectively. The fiducial scaling uses a spatial scale of $x_0 = 1 \> \rm kpc$.

\begin{table}
{\footnotesize 
\begin{tabular}{l c  }
\hline
\hline
{\bf Parameter \& units    }    & {\bf Value} \\
\hline
\hline
\multicolumn{2}{l}{\bf Equivalent Relativistic  Jet Parameters} \\
{\dag Lorentz factor}              &{5}   \\
{\dag Rest energy density/enthalpy}& {10} \\
{Velocity / Speed of light}        & {0.9798} \\
\multicolumn{2}{l}{\bf Hydrodynamic  Jet Parameters} \\
{ Pressure / External pressure }  &  {1.0} \\
{ Density  / External density  }    &  {$2.0\times 10^{-3}$}  \\
{ Mach number  }                      & {25.9} \\
{\dag Diameter} (pc)  & 40 \\
Kinetic luminosity    & {$2.77\times 10^{43} $}\\
\hline
\multicolumn{2}{l}{\bf Double isothermal potential} \\
{\dag Dark matter to } \\
{ Baryonic velocity dispersion} & {2} \\
{\dag Dark matter to } \\
{ Baryonic core radius}          & {10} \\
{ Central Baryonic density to } \\
{ Dark matter density}     & {25} \\
\multicolumn{2}{l}{\bf Dark matter values }\\
{\dag Velocity dispersion} ($\rm km \> s^{-1}$) &      400 \\
{\dag Core radius} 
(kpc)   & 3.5\\
{ Central density}  
($\rm g \> cm^{-3}$)    & $1.47\times 10^{-22}$  \\
\multicolumn{2}{l}{\bf Enclosed Masses, at $r = r_{\rm d}$ }\\
{ Dark Mass  }     ($M_\odot$)  & $8.14\times 10^{10}$  \\
{ Baryonic Mass }  ($M_\odot$) & $4.56\times 10^{10}$  \\
{ Total Mass  }    ($M_\odot$)  &  $1.27\times 10^{11}$ \\
{ Baryonic/Dark Mass ratio } &0.6 \\
\multicolumn{2}{l}{\bf Baryonic values }\\
{ Velocity dispersion} ($\rm km \> s^{-1}$) & 200\\
{ Core radius} (kpc)  &0.35 \\
{ Central density} ($\rm g \> cm^{-3}$)   &$3.68\times 10^{-21}$ \\
\multicolumn{2}{l}{\bf Enclosed Masses, at $r = r_{\rm b}$ }\\
{ Dark Mass }    ($M_\odot$)   &  $3.08\times 10^{8}$    \\
{ Baryonic Mass }   ($M_\odot$)&  $4.72\times 10^{9}$   \\
{ Total Mass }  ($M_\odot$)  &  $5.03\times 10^{9}$   \\
{  Baryonic/Dark Mass ratio } &15.3\\
\hline
\multicolumn{2}{l}{\bf Hot Atmosphere:} \\
{\dag Virial/Gas temperature} & { 1.0}  \\
Gas Temperature ($^\circ \rm K$)              & {$1.20\times 10^{7}$}  \\
\multicolumn{2}{l}{\bf Central Values:} \\
{\dag  pressure$/k$} 
($\rm cm^{-3} \> ^\circ K$)& $1.00\times 10^{6}$    \\
{ pressure} 
($\rm dynes \> cm^{-2}$)&  $1.38\times 10^{-10}$  \\
{ number density} 
($\rm cm^{-3}$)  &  $8.35\times 10^{-2}$   \\
{ mass density} 
(g cm$^{-3}$)  &  $8.64\times 10^{-26}$  \\
\hline
\multicolumn{2}{l}{\bf Warm Disk--ISM :} \\
{Virial/Gas temperature}                      & 1200.0  \\
{\dag Gas Temperature}($^\circ \rm K$)              & $1.0\times 10^{4}$ \\
{\dag Turbulent dispersion}
($\rm km \> s^{-1}$)                     &40.0 \\
{\dag Rotational Support}               &0.93 \\
\multicolumn{2}{l}{\bf Internal Non--Uniformity :} \\
{\dag Log-Normal Mean }               & 1.0    \\
{\dag Log-Normal Variance}            & 5.0     \\
{\dag Density Power Law}               & $5/3$  \\
{Volume of warm gas }
($\rm pc^3$)    & $2.55\times 10^{7}$ \\
Mass of warm gas 
($M_\odot$)       & $4.67\times 10^{5}$ \\
\multicolumn{2}{l}{\bf Central Values:} \\
{ pressure$/k$} 
($\rm cm^{-3} \> ^\circ K$)& $1.00\times 10^{6}$  \\
{\dag number density} 
($\rm cm^{-3}$)  & 10.0  \\
{ mass density} 
(g cm$^{-3}$)  &  $1.04\times 10^{-23}$ \\
\hline
\multicolumn{2}{l}{{\footnotesize Assigned parameters are indicated with a \dag symbol}} \\
\end{tabular}
\label{t:Model_A_pars}
}
\caption{Standard  Jet, Host and ISM Parameters.}
\end{table}

\section{Slice plots and surface brightness images }
\label{s:slice_plots}

We now present multi-panel snapshots from significant epochs; these snapshots are designed to bring out the relevant physics of the simulations. Some snapshots are slices of important variables such as density and pressure; in some cases we also present projected versions of variables such as the density. We also present images of the radio and X-ray surface brightness obtained by ray-tracing of the output volumes. 

In all cases the times chosen for the sequence of snapshots corresponds to the following phases in the simulation.  The mid-plane density slices (Figure~\ref{f:densslice}) most clearly illustrate the phases enumerated here, with the other representations highlighting some specific 
aspects.
\begin{enumerate}
\item Flood and Channel phase: Snapshots at 5 -- 15 kyr represent the time over which the jet
makes its way through the porous, fractal disk.  The shape of the interacting region is amorphous, and determined by flow of hot high pressure gas along weaker line in the dense disk medium.  The pressure in this phase is very high, and the x-ray emission is a strong function of time, depending on a combination of the amount of disk material that is advected and the amount of energy that is processed by radiative shocks, which increases with the size of the region.   
\item Energy Bubble phase: Snapshots at 25 -- 45 kyr represent the epoch during a high pressure bubble has grown larger than the disk,
to form a smooth nearly spherical bubble in the hot atmosphere.  The jet is still disrupted in the disk, but the bulk of the energy flux
drives the expansion of
the nearly adiabatic bubble; there is a corresponding drop in the efficiency of conversion of jet energy to thermal emission. 
 Most of the bubble grows (outside the disk) with a Radius-time power law that is consistent with classical energy driven  bubble theory.  The flood and channel behavior continues in the disk plane, but this involves a steadily decreasing fraction of the jet energy flux.
\item Jet Breakout phase: In the 55 -- 70 kyr epoch the jet breaks free of the few remaining clouds in its path and the jet terminus propagates towards the edge of the bubble. The Bubble has a low density, so that the jet transits the bubble quickly.
\item Classical Phase: At  $t \approx 75 \> \rm kyr$ the jet pierces the original bubble and subsequently forms a classical radio lobe with hotspot, cocoon, bow-shock and backflow.  The remnant of the spherical bubble continues to grow, and flow continues within the disk, with numerous radiative and non-radiative shocks throughout the whole disk persisting to late times excepting the very centre of the disk, which has been cleared by the main jet.
\end{enumerate}

\subsection{Mid-plane Density and Pressure Slices}
\label{s:density}

\begin{figure}
\begin{center}
\includegraphics[width=5in]{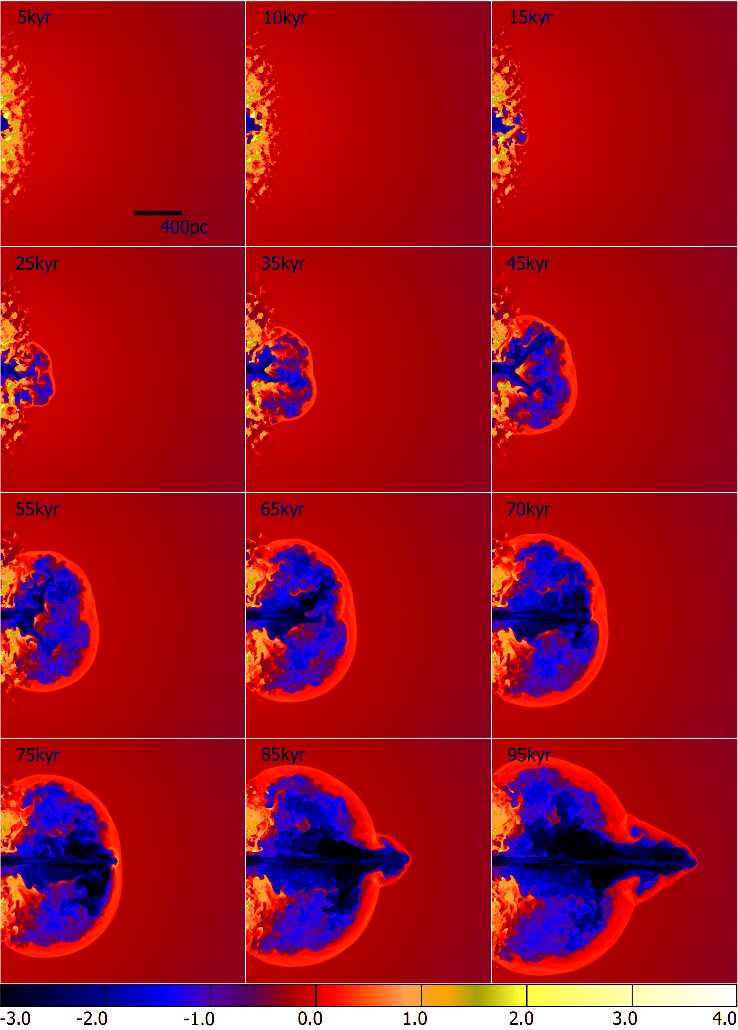}
\caption{\footnotesize Mid-plane density slices: The panels represent the logarithm of the density in a $k=255$ plane of the $512^3$ simulation in the various phases of the evolution described. The grayscale bar shows the range of the density. The selection of snapshot times is the same in all subsequent figures.}
\label{f:densslice}
\end{center}
\end{figure}

Figure~\ref{f:densslice} shows a sequence of density slices.
At 10 and 15 kyr ``bright'' spots of high density, caused by radiative shocks driven into the warm gas, are evident. The jet is anything but straight as it selects a path of least resistance through the region in which the initial density is  described in terms of a log-normal distribution (see \S~{s:lognormal}).

In the (25 -- 45 kyr) phase a dense obstruction causes the jet to split. Further regions of high density are observed, often at the tips of dense cloud material that is being ablated by the outflowing radio plasma. The thermalized jet pressure drives an almost spherical bubble in the surrounding medium. A bow-shock surrounds the bubble and the contact discontinuity between the bubble and the hot interstellar medium is apparent.

In the jet-breakout phase (55 -- 70~kyr) the pressure of shocked jet plasma, continues to drive a more nearly spherical bubble into the hot interstellar medium. Towards the end of this phase, the obstructing cloud that was responsible for the disruption of the jet dissipates and a straight jet channel is starting to be established. 

At 75~kyr (the lowest left panel of Figure~\ref{f:densslice}) a unidirectional jet flow has been established and the jet is about to pierce the initial bubble. Subsequently, the jet propagates beyond the bubble and starts to establish a classic FR2 lobe morphology consisting of bow-shock, cocoon  and backflow.   
Another way of looking at this phase is that the quasi-isotropic (energy-driven) phase of radio lobe expansion makes a transition to a bow-shock dominated (momentum-driven) phase as the now relatively undeflected jet progresses through the interstellar medium. An interesting feature, which is discussed further below, is that the jet, whilst maintaining a more or less single direction, is unstable. As a result of both filamentation and helical instabilities, the end of the jet has become somewhat diffuse and the jet momentum is spread over a region which is wider than the original jet diameter in a similar manner to that envisaged by \citet{scheuer82a}.

\begin{figure}
\begin{center}
\includegraphics[width=5in]{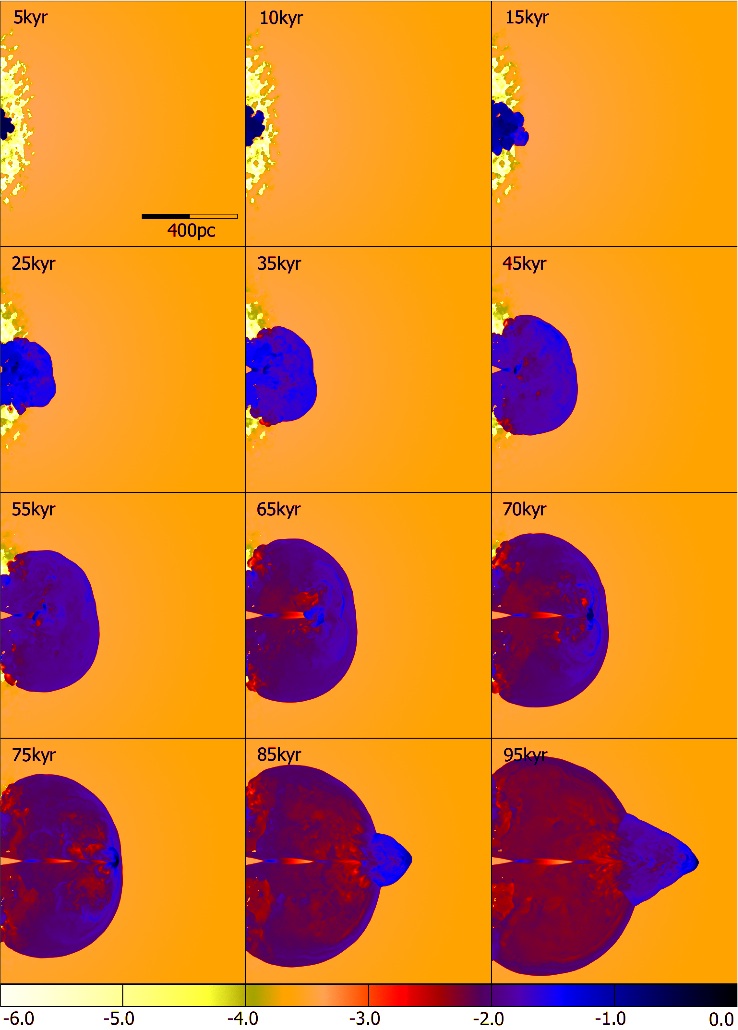}
\caption{\footnotesize Mid-plane pressure slices: The panels represent the logarithm of the pressure in an $k=255$ plane of the $512^3$ simulation at the same phases of the evolution as in Figure~\ref{f:densslice}. Note that for clarity, the sense of the grayscale colorbar has been reversed compared to that for the density.} 
\label{f:presslice}
\end{center}
\end{figure}

Mid-plane pressure slices in Figure~\ref{f:presslice} confirm the evolutionary features revealed by the density and at the same time show some additional characteristics. 

The disk interaction phase shows a high-pressured amorphous region where the jet struggles to make its way out of the confining disk. The lower pressure dense gas (the white region) surrounds this amorphous bubble.  In the flood and channel phase the pressure in the jet-driven bubble is fairly uniform (as expected) but also shows at 35 kyr a biconical shock associated with the compression of the emerging jet by the overpressured bubble together with a bow shock caused by obstructing high density gas in the path of the jet. There are also some spots of high pressure associated with the general turbulence in the bubble.  During the jet breakout phase most of the radio lobe is at constant pressure, and there are only small residual regions of low pressure associated with the disk. Further biconical shocks form within the jet and the high pressure jet terminus is seen propagating through the bubble.  During the classical phase the hot spot at the end of the jet pierces the bubble and forms a higher pressure, momentum-driven cocoon. The biconical shocks do not persist within this cocoon and the high pressure region at the end of the jet is spread out over more than a jet diameter. 

\subsection{Radio Surface Brightness}
\label{s:inu}

We construct a surface brightness emissivity from the pressure $p_{\rm nt}$ of non-thermal plasma, defined to be that which originates from the jet. The emissivity is defined by:
$ j_\nu \propto p_{\rm nt}^{(3+\alpha)/2} \, \nu^{=\alpha}$,
where $p_{\rm nt}$ is the nonthermal pressure and $\alpha$ is the spectral index (with the convention $F_\nu \propto \nu^{-\alpha}$). This expression assumes that the magnetic pressure tracks the plasma pressure, that is, $B^2/8 \pi \propto p$.  The scalar tracer $phi$, which is the mass concentration by density of the non-thermal plasma, is used to indicate the presence of jet plasma at the various regions of the grid.  We use the above expression  for the emissivity when $\phi > 10^{-3}$.

\begin{figure}
\begin{center}
\includegraphics[width=5in]{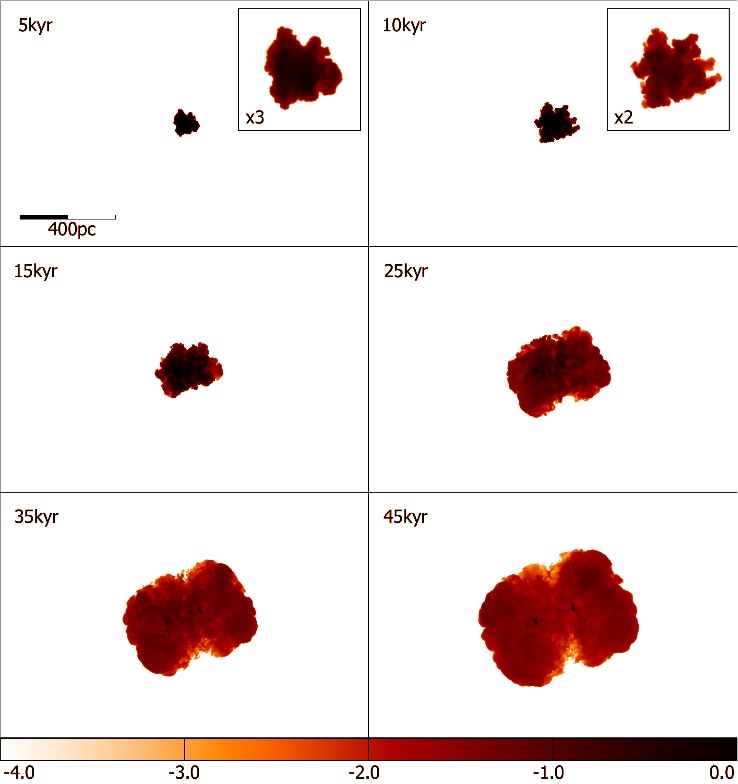}
\caption{\footnotesize Radio Surface Brightness: The panels represent the logarithm of the synthetic radio  surface brightness at the 5 -- 45~kyr phases of the evolution, corresponding to the first 6 panels of the density evolution in Figure~\ref{f:densslice}. The insets for 5~kyr and 10~kyr show the surface brightness images magnified by factors of 3 and 2 respectively in order to bring out the detail in these early stages.}
\label{f:inu_A}
\end{center}
\end{figure}

\begin{figure}
\begin{center}
\includegraphics[width=5in]{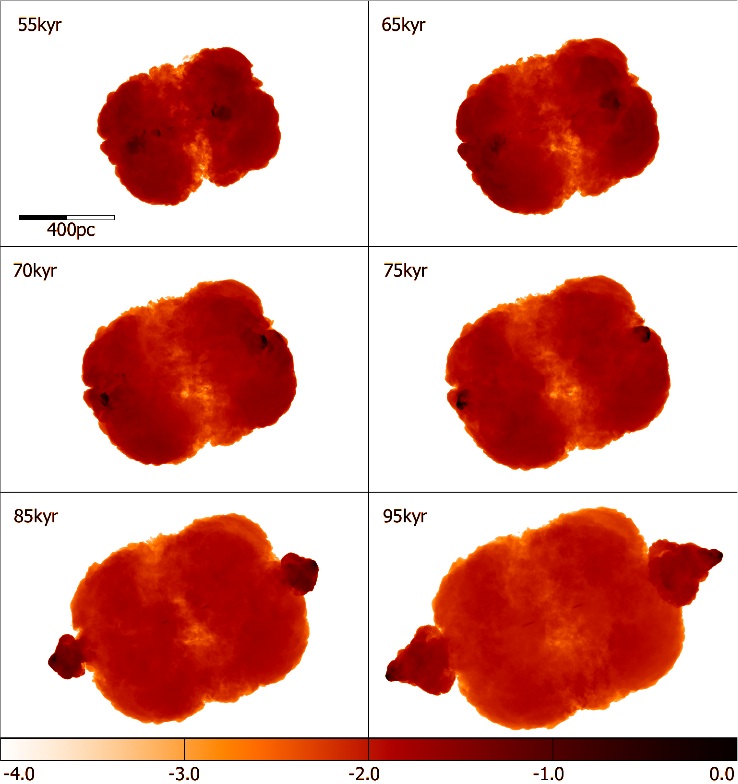}
\caption{\footnotesize Radio Surface Brightness: The panels represent the logarithm of the synthetic radio  surface brightness at the 55 -- 95~kyr phases of the evolution, as in the second 6 panels of Figure~\ref{f:densslice}.}
\label{f:inu_B}
\end{center}
\end{figure}

A series of synthetic radio surface brightness snapshots is displayed in Figures~\ref{f:inu_A} and \ref{f:inu_B} at the same times as in the previous figures.  At the early times the surface brightness shows the amorphous structure that is evident in the density slices discussed previously.

In the flood and channel phase two radio-emitting bubbles form. There is some flocculent structure in the surface brightness resulting from the non-uniform thermalisation of jet plasma during this phase. We also see quite clearly (particularly at 35~kyr and 45~kyr) the bow shock (on each side) caused by the impact of the jet on the obstructing cloud discussed in \S~\ref{s:density}.

During the jet-breakout phase  the radio-emitting bubbles become smoother in appearance but high surface brightness features associated initially with the jet-cloud bow shock (55~kyr) and then with the jet terminal shock (65 and 70~kyr) are also seen propagating out through the respective bubbles. Another feature which becomes obvious here, but which is also evident from approximately 25~kyr onwards, is a band of low surface brightness emission caused by the partial exclusion of radio-emitting plasma from the disk region. 

From 75~kyr onwards we see the jets breaking free of the bubbles.

\subsection{X-ray emission}
\label{s:xray}

\begin{figure}
\includegraphics[width=5in]{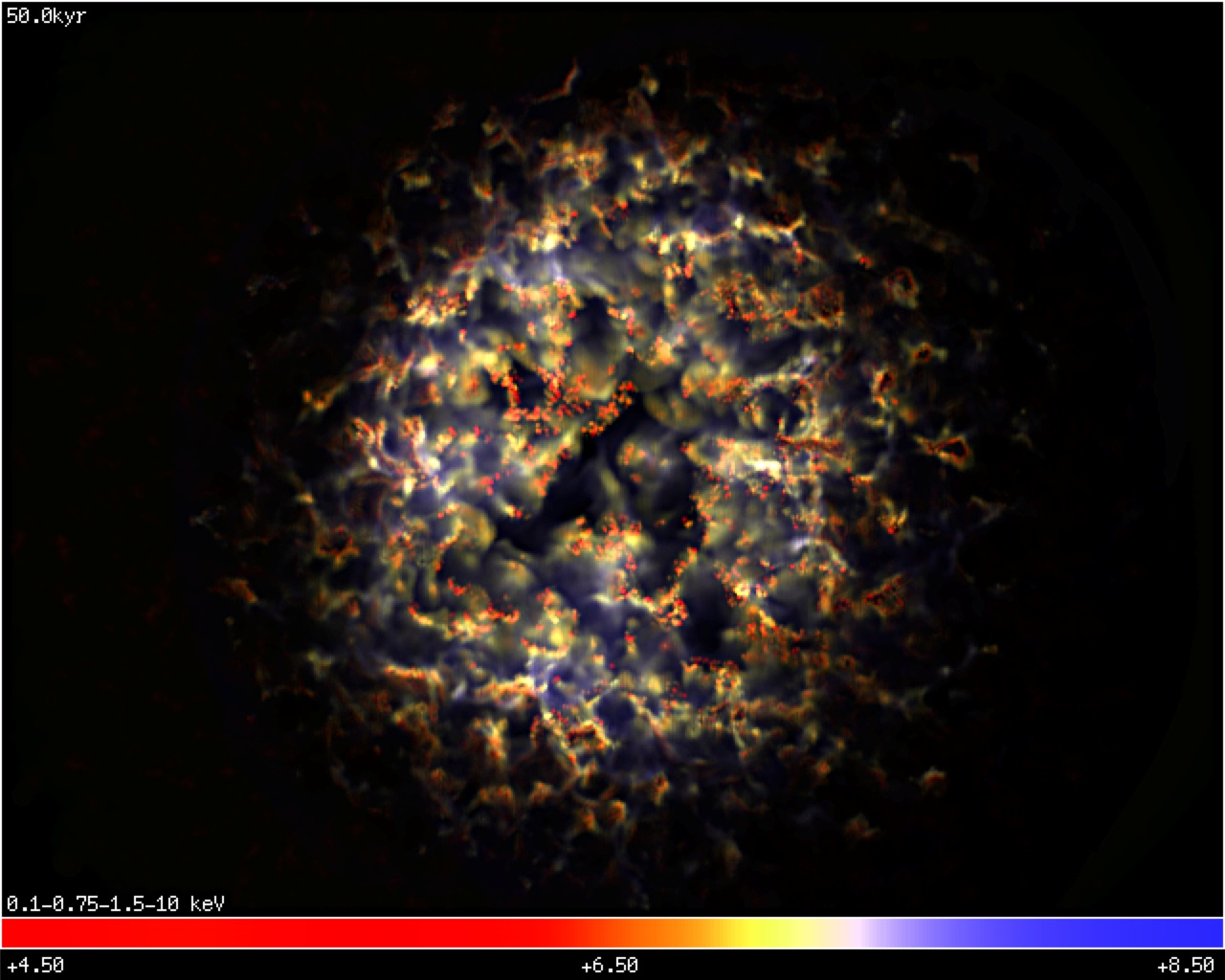}
\caption{\footnotesize  Soft X-rays: the logarithm of the soft X-ray surface brightness nearly face on at 50 kyr.  Red indicates emission between 0.1 and 0.75 keV, and marks the location of slow, fully radiative shocks.  Green represents emission between 0.75 and 1.5 keV, and comes from gas $> 10^7$~K, while blue comes from emission betwen 1.5 and 10 keV, and shows how weakly the very hottest gas ($> 10^9$~K) is cooling.}
\label{f:xsoft}
\end{figure}

\begin{figure}
\includegraphics[width=5in]{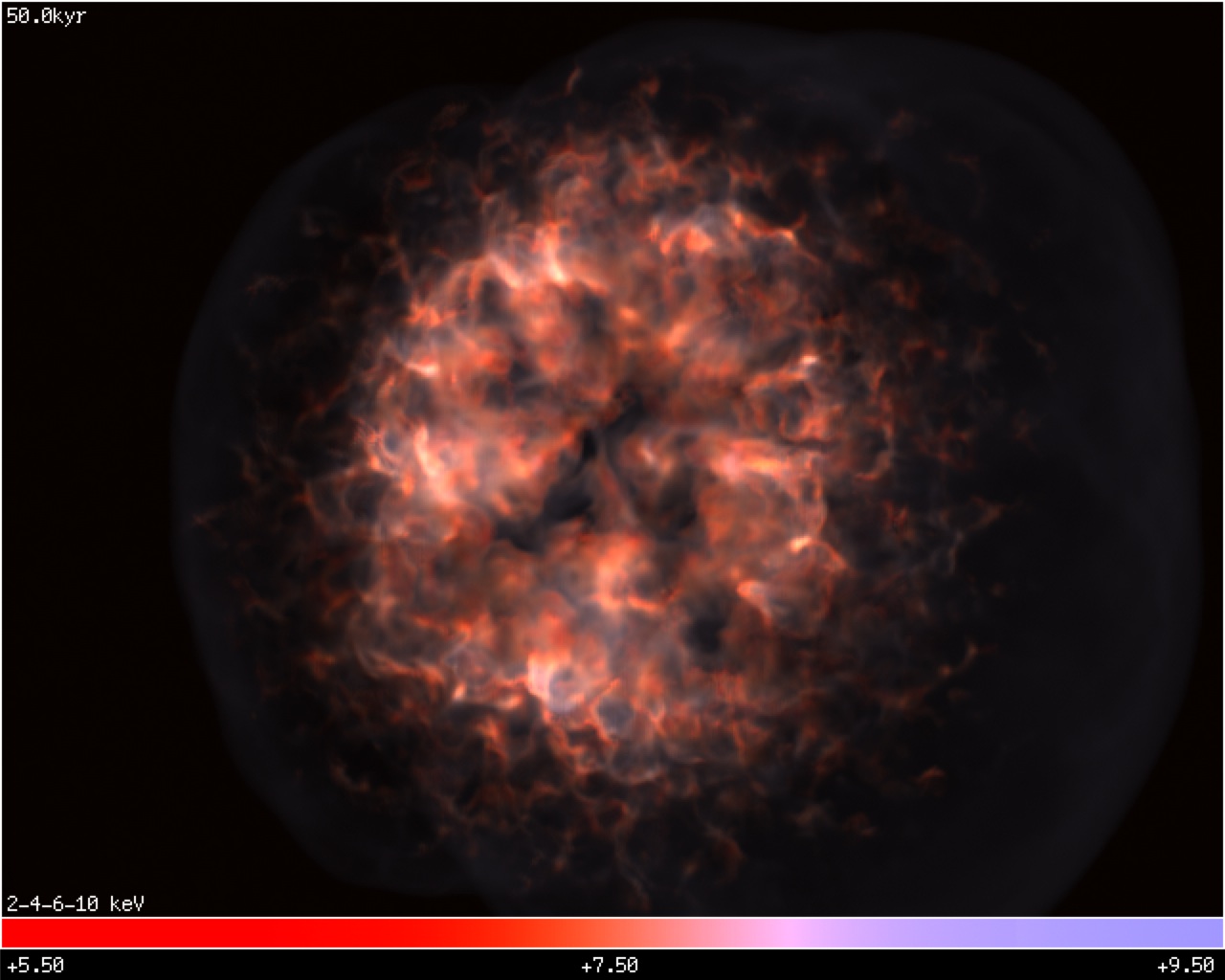}
\caption{\footnotesize Hard X-rays: as for the previous figure, but for the hard X-ray bands Red = 2--4, Green = 4--6 and Blue = 6--10!keV emission.  The very hot outer bubble shock is just visible in the hardest bands, and the red emission shows gas up to $\sim 3 \times 10^7$~K being advected around the dense disk clouds. }
\label{f:xhard}
\end{figure}

Images of the  X-ray surface brightness present a different view of the physics of this simulation. The temperature of this gas and the total thermal cooling per unit volume, $\rho^2 \Lambda_\rho(T)$, determine the spectrum of gas at each voxel, via the original MAPPINGS spectrum used to define the cooling function. The spectrum is integrated 
over specific bands to determine X-ray emissivities in those bands. These emissivities are then integrated along the line of sight to determine the X-ray surface brightnesses. With three bands, false colour RGB images can be constructed and these are presented in Figures~\ref{f:xsoft}, and \ref{f:xhard}.

The soft X-ray image,  Figures~\ref{f:xsoft}, seen nearly face on to the disk at $t = 50$~kyr, uses bands between 0.1 and 1.5 keV for red and green components, with the hard 1.5-10keV component in (faint) blue. This view shows how the dense clouds show sub-structure in the X-rays, where red dots pick out slow fully radiative shockwaves at the limit of the simulation resolution.  The higher energy green and blue emisison traces the hotter non-raditatively shocked ISM gas that is subsequently advected along with the hot bubble.  This radiative/non-radiative divide separates inefficient and efficient mass transport, and a significant fraction of the densest gas is radiatively shocked and only inefficiently transported by the expanding bubble. 
 Figures~\ref{f:xhard} suggests that the majority of the advected gas and the hot bubble itself is primarily visible in in emission above 2~keV.
 
\begin{figure}
\includegraphics[width=5in]{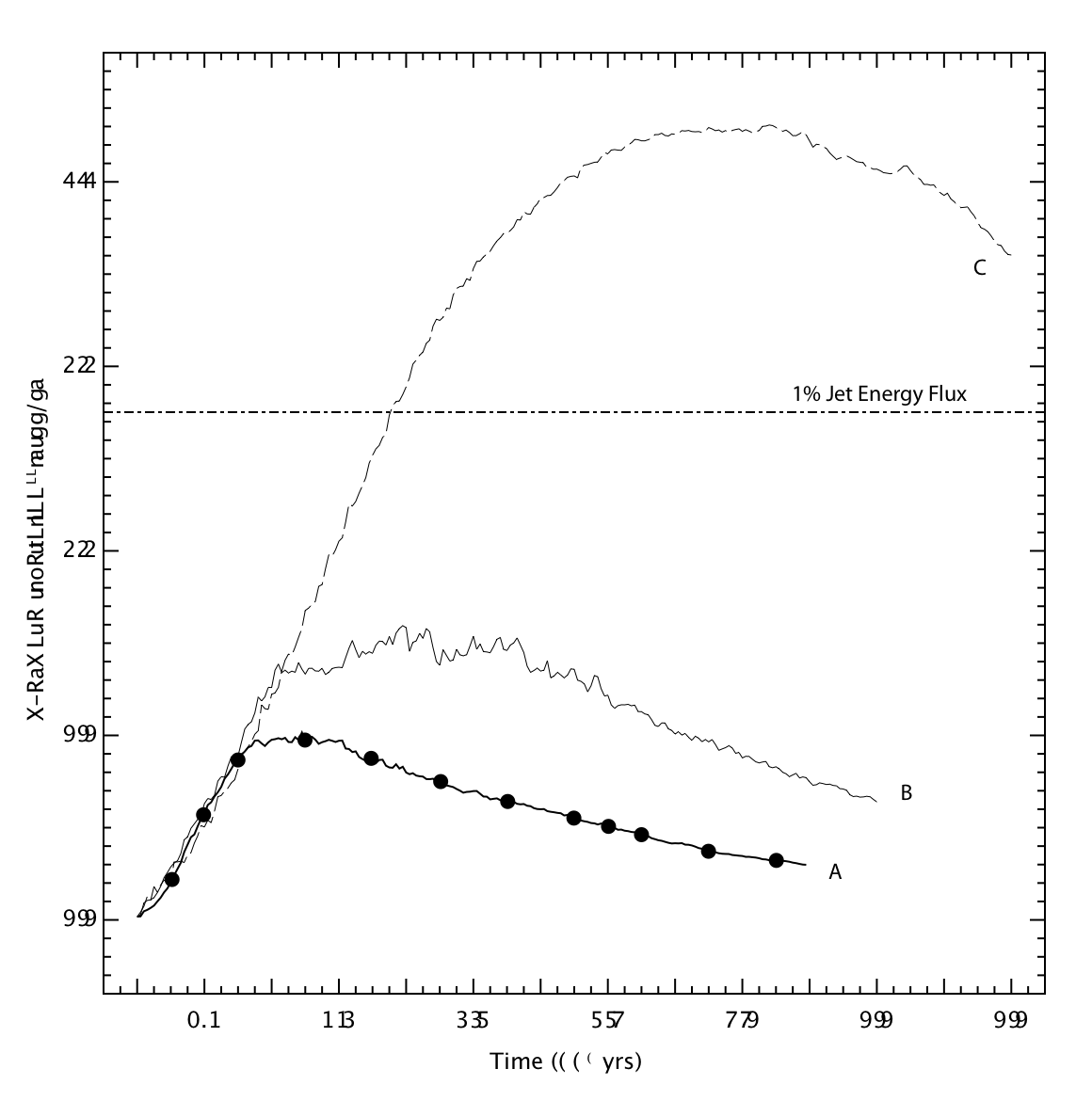}
\caption{\footnotesize The evolution of the $0.1-10.0 \> \rm keV$ X-ray luminosity of models {\bf A} , {\bf B}  and {\bf C}. The times of the snapshots used in the detailed description of model {\bf A} are indicated by filled circles. A line indicating 1\% of the jet energy flux is also drawn.}
\label{f:L_X}
\end{figure}

The total $0.1 - 1.0 \> \rm keV$ X-ray emission for models {\bf A}, {\bf B}  and {\bf C} as a function of time is shown in Figure~\ref{f:L_X}. The common feature of all the luminosity curves is that they rise approximately linearly to a broad maximum and then decline. For the more realistic models  {\bf A} and  {\bf B}, the maximum luminosity is a fraction  of one percent of the jet energy flux. In the physically unrealistic model~ {\bf C}, the peak X-ray luminosity is about 1.6\% of the jet energy flux.    In all cases, the peak X-ray luminosity occurs when the jet can be considered to have just broken free of the disk. Before that epoch a portion of the jet power is being directed into the radiative shocks that are responsible for the X-ray emission. When the jet breaks out, the fraction of its power that was being diverted into radiative shocks is reduced since the jet power is now diverted into non-radiative shocks in the hot interstellar medium.

\section{Mass Redistribution}

One of the main motivations for examining jet--ISM interactions in detail is to examine the importance of black hole induced feedback processes in galaxy formation. Here we consider the mass that is driven to larger scales as a result of the interaction of the jet with the turbulent disk.

In addition to the main simulation model~{\bf A}, models {\bf B} and model~{\bf C} were performed to look for changes in the interaction sequence with a change in two disk ISM parameters, density and uniformity.
 Model B differs only in the mean density of warm gas. 
The central value in model {\bf B} is $20 \> \rm cm^{-3}$ compared to $10 \> \rm cm^{-3}$ in model~{\bf A}. The point of model {\bf B} is to examine the effect of the mean density of identically distributed gas. Model {\bf C}, has the same mean density as model {\bf B}, but has a perfectly smooth distribution of warm gas. The point of this model is to examine the effect of the porosity of the gas distribution. 

The resolution of Models {\bf B} and {\bf C} is a factor of two lower, {\em i.e.} $256 \times 256 \times 256$. However, this does not appear to affect the comparison; similar behaviour is observed and the jet is adequately resolved with 10 resolution elements across its diameter. We do not expect the much more uniform density distribution in model {\bf C} to suffer from lower resolution. 

\begin{figure}
\begin{center}
\includegraphics[width=5in]{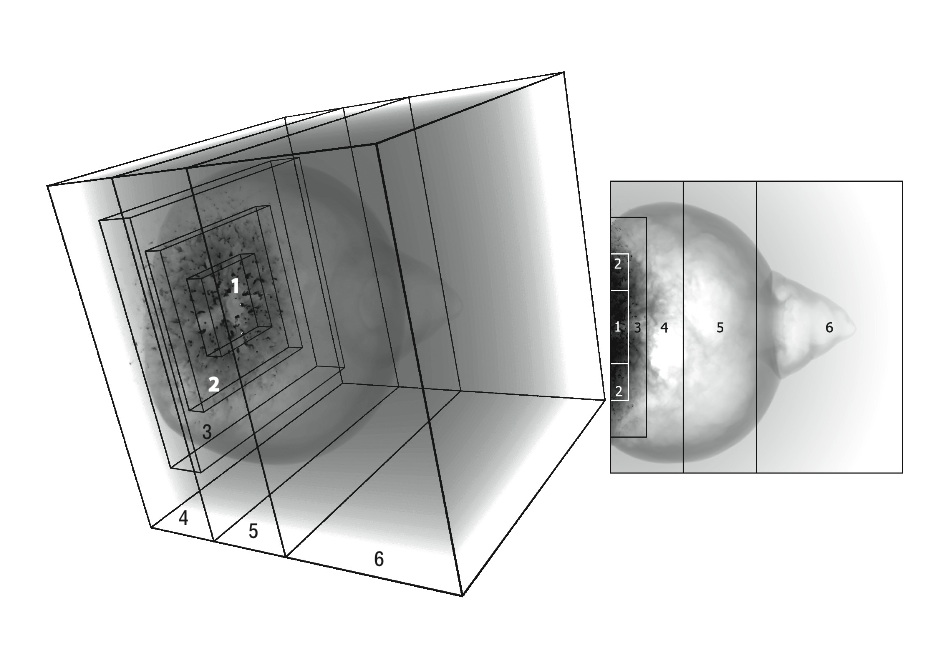}
\end{center}
\caption{{\footnotesize A 3D perspective and side view of the arrangement of the regions in the mass integration analysis.  Each zone encompasses the preceding zones, each being a rectangular integration box, and the labeled regions are defined at the difference between each successive pair of zones, with region 1 and zone 1 being the same.} }
\label{f:masszones}
\end{figure}

 We have divided the computational domain into a nested series  of rectangular sub-zones, and define integration regions as the difference between successive zones as follows.   Integration of the density in each zone give the zone masses as a function of time, and the differences between successive zone integrals allow us to monitor the transport of material from region to region.  The zones are rectangular for integration simplicity, although the regions between them may not be.  Figure \ref{f:masszones} shows the regions graphically for a late time density snapshot of Model~{\bf A}.

\begin{figure}
\begin{center}
\includegraphics[width=5in]{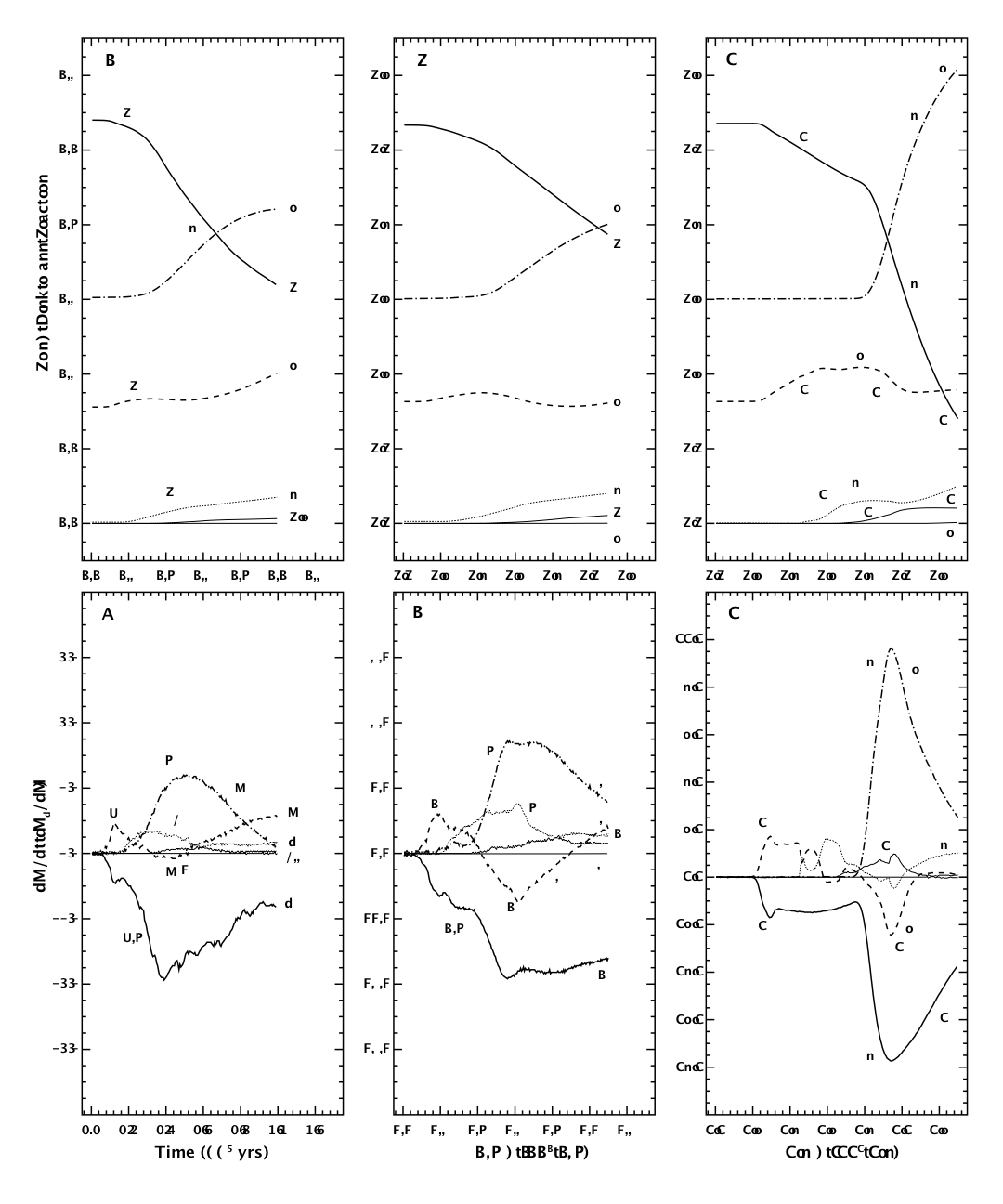}
\caption{{\footnotesize  Mass fractions and derivatives in regions 1-6, for model {\bf A} (left), {\bf B} (center) and {\bf C} (right).  The upper panels show the fraction of disk material, by mass ,in each of the 6 regions, labelled by region number.  The letters {\bf U}, {\bf P} and {\bf F} indicate the dominant direction of mass transfer on the curves. {\bf U} indicate periods of '`uplift' .  {\bf P} indicate in plane movement.  {\bf F} indicates where material is predominately falling back to a lower zone. }}
\label{f:mma}
\end{center}
\end{figure}

\begin{figure}
\begin{center}
\includegraphics[width=5in]{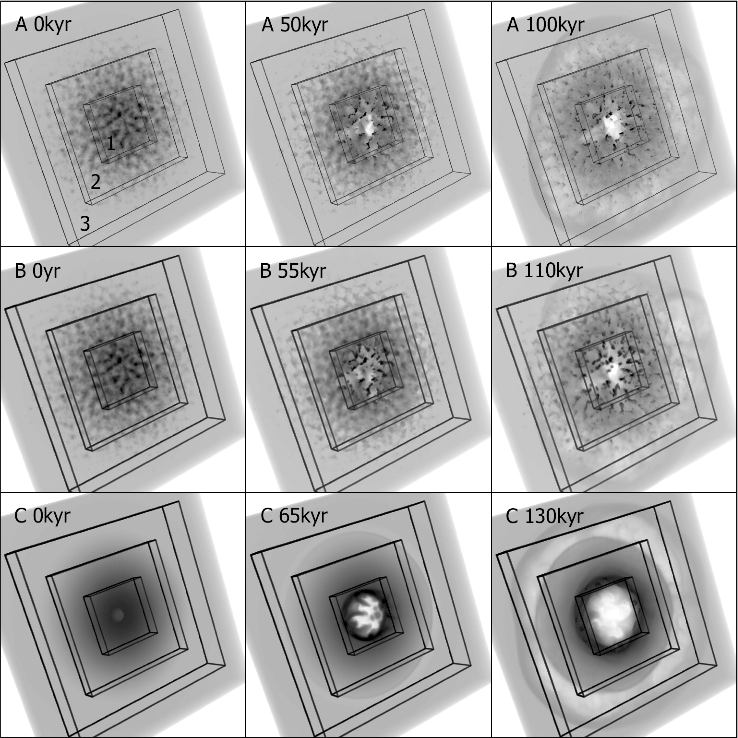}
\end{center}
\caption{{\footnotesize  These images show the distribution of gas density at the beginning, midpoint and final state for the three models in zone 4, {\em i.e.} just the lower 256~pc ($1/4$) of the grid, excluding the upper atmosphere regions 5 and 6 for clarity.  The endpoints were determined by halting the simulation once the jet reached the far edge of the computational grid.  The location of regions 1-3 are shown.}}
\label{f:abc_zones}
\end{figure}

Mass redistribution in the simulations is summarised in  Figures~\ref{f:mma} and \ref{f:abc_zones}. The upper panels in Figure~\ref{f:mma} show, for each model, the disk mass fraction, defined as the fraction of mass in each region occupied by matter originally in the warm disk, as a function of time. The lower panels show the time rate of change of mass in each region as a function of time.   Region 3 shows a range of processes, uplift from zone 1 and fallback to zones 1 and 2, and the mass in region 3 is variable.  The upper atmosphere regions, 5 and 6, see very little of the disk material by the end of the simulations, and even with significant movement of the disk in model {bf C}, 2\% or less ends up above 500~pc from the disk plane.   The mass exchange rates in models {\bf A} and {\bf B} are similar, peaking at around 2 M$_\odot$ per year in the plane of the disk between region 1 and 2, while the monolithic disk shock in the uniform test model {\bf C} generates rates up to 5 times greater.  
Models {\bf B} and {\bf C} are identical except for the non-uniformity of {\bf B} suggesting that quantitative models of mass transfer in jet-host feedback models needs to take non-uniformity into account. Detailed knowledge of the non-uniformity in real radio galaxy hosts is needed before truly realistic transport models can be computed.


Figure~\ref{f:abc_zones} shows a projection of the density in the first 3 regions at selected epochs.  It illustrates how the single disk shock in the uniform model, {\bf C}, is effectively sweeping the inner regions of the disk clear of high density gas, in contrast to the non-uniform models {\bf A} and {\bf B}.

Three key points about  the mass redistribution are evident from the figures:
\begin{enumerate}
\item {\bf Mass Advection Efficiency with a Radiative Critical Density}
The initial and final mass fractions in  Figure \ref{f:mma}, show that the two clumpy models ({\bf A} and {\bf B}) are similar, with a significant fraction of disk material remaining in the inner region~1.  The relative mass remaining in Model {\bf B} is somewhat greater and {\bf A}. This may be consistent with the notion of a critical density,  
$\rho_{\rm cr}$, above which the traversing shocks become locally fully radiative, cool and compress gas efficiently forming small dense, slow -moving clumps. For $\rho < \rho_{\rm crit}$ the shocks may be adiabatic, heating and expanding the gas which becomes advected rapidly along with the hot gas of the main blast.

If there is such a critical density $\rho_{\rm cr}$, then we expect that in the denser Model~{\bf B} the mass {\em fraction} compressed into dense clumps would be higher since $\rho_{\rm cr}$ is determined by the jet ram pressure and the cooling function, and the corresponding fraction above a fixed point in the denser model is greater.    
In region 1 in Model~{\bf A}, about 60\% of the original mass remains at late times, and in model~{\bf B} at a similar late dynamical time (estimated by the extent of the main jet-bubble),  72\% remains. This suggests that as long as the material above a radiative critical density remains in dense, un-connected, clumps, the non-gaussian log-normal statistics of the density may influence the behavior of the global system. 

\item {\bf Inefficient Transport of Disk ISM to outer Hot Atmosphere}
In all cases, about 83--85\% of the gas starts in zones 1 and 2 combined, and 75--78\% remains in the two zones at the end of the simulation, with only about 4--7\% of the disk gas being transported to the higher regions 4, 5 and 6.  Only trace amounts of disk material are transported to region~6 and this is largely in the form of material diffused into the very hot radio plasma. This fraction is uncertain since it may be dominated by numerical diffusion rather than physical transport.  Despite the fact that the main blastwave crosses the entire disk, and locally {\em many} shock crossing timescales pass, the majority of the warm clumpy disk ISM is not moved out of the plane of the disk, and the jet outburst is inefficient  at `clearing out' the disk.

\item {\bf Transport of Disk ISM within the Disk}
In the uniform model~{\bf C}, 75\% of the disk mass initially in region 1 is swept to region 2, and as the circular empty region increases we expect that essentially all of the inner disk would be swept clear in another $20-40$~kyr.  In models {\bf A} and {\bf B}  only 30--40\% of the disk in region~1 is transported to region~2, and the timescale for clearing out region ~1 is becoming greater than 150,000 years (assuming constancy of the final mass loss rates).  It is therefore likely that dense material will remain in region ~1 for a longer time in models {\bf A} and {\bf B} compared to model {\bf C}.  
\end{enumerate}

The clumpy nature of the models {\bf A} and {\bf B} significantly extend the lifetime of material in the inner zones of the outbursts, showing that in the cirumnuclear region of an active galactic nucleus, the inhomogenity of gas in the inner region makes it possible to retain gas there, maintaining a supply of accreting gas to the black hole for longer than models with homogeneous gas distributions may suggest. The relevant timescale for the duration of the outburst is determined primarily by  the accretion timescale. 

\section{Summary}

The main simulation that we have presented, of a jet interacting with a turbulently supported disk, exhibits four interesting phases: 1) A `flood and channel'  phase 2) A nearly adiabatic (globally) energy driven bubble phase 
3) A jet breakout phase  4) A classical phase. These are described in detail in \S~\ref{s:slice_plots}.

The late-time morphology of the radio source manifests a clear signature of the two earlier phases. The X-ray emission also presents some further insights. Not only do we see X-ray emission from the disk associated with radiative shocks in the early stages and X-ray emission associated with the bow-shock of the radio lobes but we also see persistent X-ray emission from the disk well after the radio bubble has passed through the disk. This is caused by the continued driving of radiative shocks into the disk by the high pressured bubble. The peak in the X-ray luminosity occurs during the jet breakout phase, when more of the jet energy flux is diverted into the growing, hot adiabatic bubble and there is less available for processing in the dense disk material.  The behaviour of the disk is suggestive of a critical phenomenon: The critical cloud density at which the ram pressure of the jet is converted into fully radiative shocks, and the mass fraction that is available for rapid mass advection may be influenced by the log--normal density field in the disk.  Additionally , the global movement of disk material throughout the simulation is clearly very different for a uniform disk model compared otherwise similar, albeit somewhat ad hoc, non-uniform models.

\begin{acknowledgements}
We thank the ANU supercomputing facility time assignment committee for allocations of computer time. This research was also funded by ARC discovery project grants DP0345983 and DPDP0664434.
\end{acknowledgements}

\end{document}